\documentclass[slac_one]{revtex4}
\usepackage{bm}
\usepackage{subfigure}
\usepackage{graphicx}
\usepackage{fancyhdr}
\def\ppbar{\mbox{p}\overline{\mbox{p}}}

\def\MET{{E\kern -0.6em/_{\rm T}}}
\def\METx{{E\kern -0.6em/_{\rm x}}}
\def\METy{{E\kern -0.6em/_{\rm y}}}
\def\PET{{P\kern -0.6em/_{\rm T}}}

\def\Mvis{{M_{\rm vis}}}

\begin{document}

\title{Search for MSSM Higgs Boson Production in Di-tau Final States with $\mathcal{L}=2.2 fb^{-1}$  
at the D\O\ Detector}

\author{ Wan-Ching Yang on behalf of the D\O\ Collaboration}
\affiliation{School of Physics and Astronomy, \\
The University of Manchester, United Kingdom}

\begin{abstract}
A search for the production of neutral Higgs bosons decaying
into $\tau^+\tau^-$ final states is presented in $p$$\bar{p}$ collisions
at a center-of-mass energy of 1.96 TeV.  
The integrated luminosity used for the study is about $2.2$~fb$^{-1}$, collected by the D\O\ Experiment 
at the Fermilab Tevatron Collider. No significant excess is observed over the background expectation.
The results are interpreted in the Minimal
Supersymmetric Standard Model (MSSM) and regions in the ($m_{A}$, tan$\beta$) parameter space for two MSSM benchmark
scenarios are excluded.

\end{abstract}

\maketitle

\thispagestyle{fancy}

\section{\label{sec:intro}INTRODUCTION}

Neutral Higgs bosons produced in $\ppbar$ collisions at the Tevatron 
can decay into $\tau^+\tau^-$ final states.  The cross-section times branching ratio
of the $h\rightarrow\tau^+\tau^-$ final state in the Standard Model (SM) 
is too small to play any role in SM Higgs boson searches
due to the large irreducible background from Drell-Yan production
in the interesting (low mass) region. This, however, is different in the 
Minimal Supersymmetric Standard Model (MSSM), which predicts
two Higgs doublets leading to five Higgs bosons: a pair of charged Higgs boson (H$^{\pm}$); two neutral
CP-even Higgs bosons (h,H) and a CP-odd Higgs boson (A). 
At tree level, the Higgs sector of the MSSM is fully described
by two parameters, which are chosen to be the mass of the CP-odd
Higgs boson, $m_A$, and $\tan \beta$, the ratio of the vacuum expectation
values of the two Higgs doublets. The Higgs boson production cross-section
is enhanced in the region of low $m_A$ and high $\tan\beta$ due
to the enhanced Higgs boson coupling to down-type fermions~\cite{mssm-benchmark}.
In addition, two of the three neutral Higgs bosons, commonly denoted
by $\phi$, are often nearly degenerate in mass, leading to a further increase in the cross-section. 
In the low $m_A$, high $\tan\beta$ region of the parameter space, 
Tevatron searches can therefore probe several MSSM benchmark scenarios extending the search regions
covered by LEP~\cite{bib-lep}. 

Searches for neutral Higgs bosons decaying into tau lepton pairs, $\phi$ $(=H,h,A)\to\tau\tau$, 
have been performed by the D\O\ Collaboration with integrated luminosities of $\mathcal{L}=1.0$~fb$^{-1}$ in Run IIa~\cite{bib-d0-a} and
$\mathcal{L}=1.2$~fb$^{-1}$ in Run IIb~\cite{bib-d0-b}.  The Run IIa search requires the tau pairs 
to decay into $\tau_{e}\tau_{had}$, $\tau_{\mu}\tau_{had}$, or $\tau_{e}\tau_{\mu}$, and the Run IIb search
requires the tau pairs to decay into $\tau_{\mu}$$\tau_{had}$, where $\tau_{e}$ and $\tau_{\mu}$ 
are the leptonic decays of the tau and $\tau_{had}$ is the hadronic decay mode. 
These studies together represent a data set of $\mathcal{L}=2.2$~fb$^{-1}$.

The search strategy relies primarily on 
implementing an efficient tau identification algorithm in conjunction to a series of selections that 
remove backgrounds, which are dominated by electroweak $Z/\gamma^{*} \rightarrow \tau\tau$ and 
$Z/\gamma^{*} \rightarrow \mu\mu$ processes, as well as those from heavy-flavor multijet 
events where a jet can be misidentified as a $\tau$ candidate.

\section{FINAL RESULTS}

The visible mass, $\Mvis$, is used to search for the signal in the data sample.
This variable is defined as:

\begin{equation}
\Mvis  =  \sqrt{(P_{\tau_1}\ + \ P_{\tau_2} \ + \ \MET)^2},
\end{equation}

\noindent
and is calculated using the four vectors of the visible tau decay products, $P_{\tau}$, and
of the missing momentum $\PET = (\MET,\METx,\METy, 0)$.
$\METx$ and $\METy$ indicate the components of $\MET$.
No significant excess is seen over the background expectation 
,and thus limits on the production cross section for neutral Higgs boson times the branching 
fraction into tau leptons are given for neutral Higgs bosons in the mass range 90 to $300~GeV$. 

\begin{figure}[htbp]
   \begin{center}
     \subfigure[] {
       \includegraphics[width=7cm]{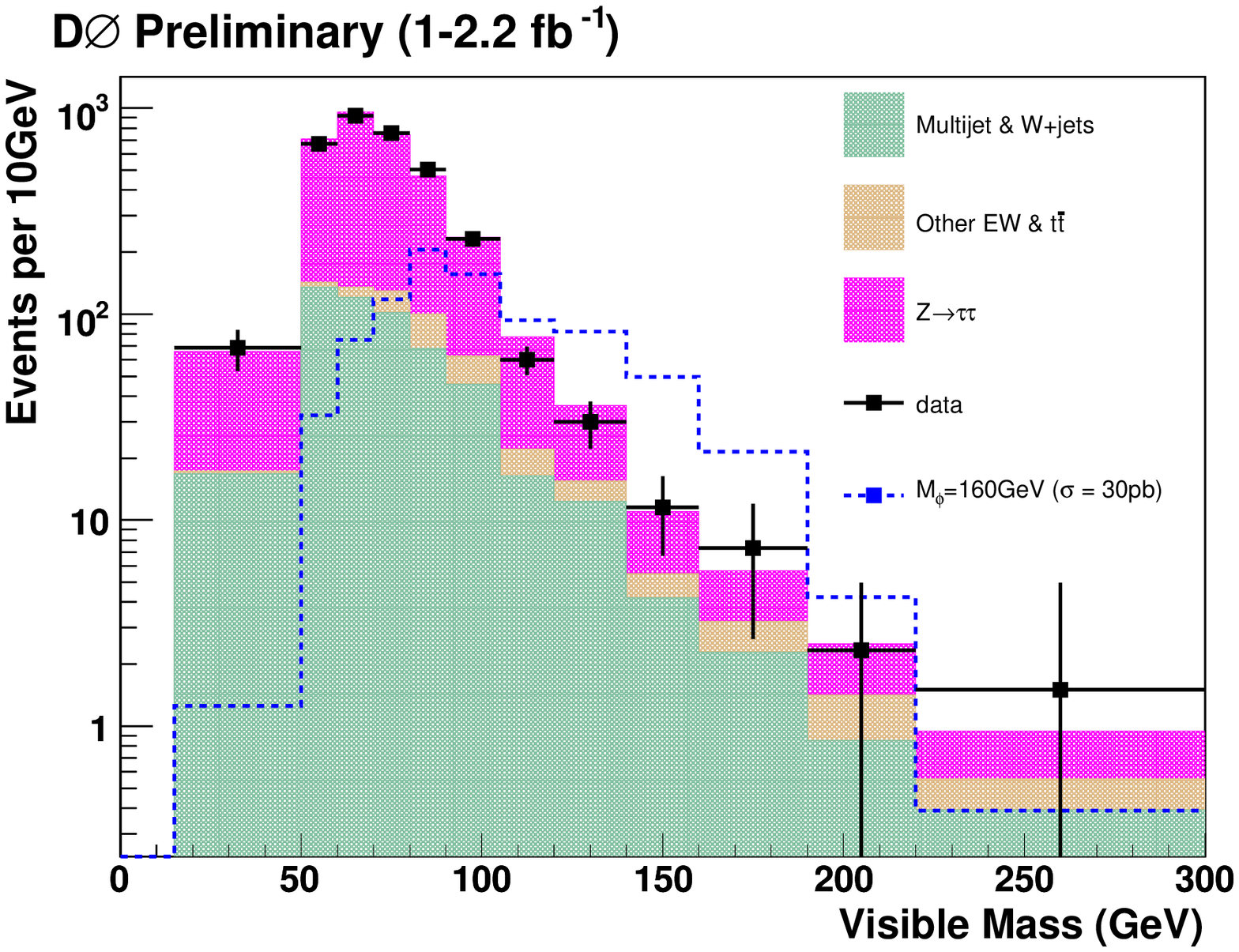}}
     \subfigure[] {
       \includegraphics[width=7cm]{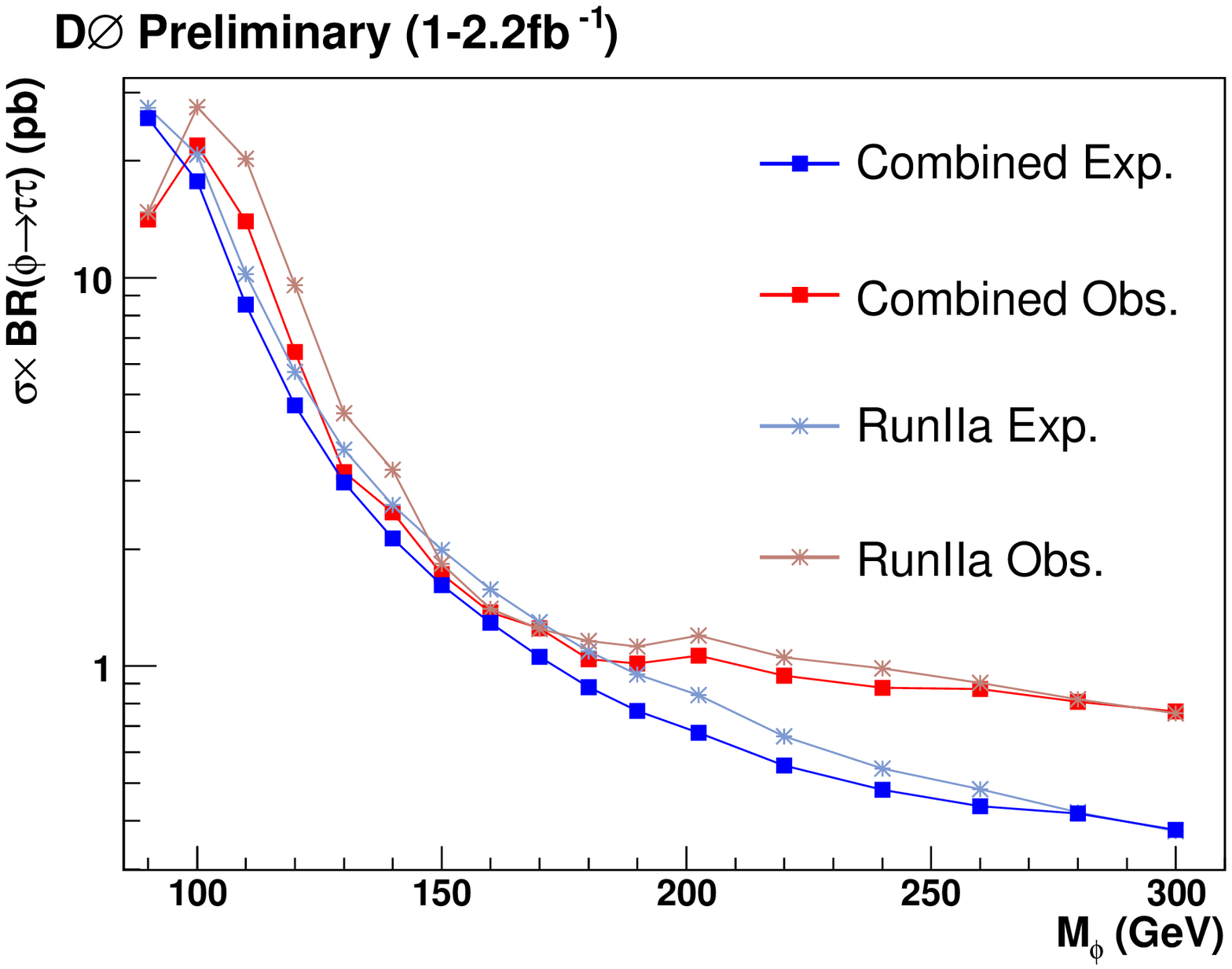}}
     \end{center}
   \caption{
     (a) Distribution of the visible mass after all selections. 
     The data, shown with error bars, are compared to the sum of the expected backgrounds.
     Also shown, in blue open histogram, is the signal for a Higgs mass of 160 GeV assuming a signal cross section times branching fraction of 30 pb.
     (b) Expected and observed upper limits on the production cross section times branching ratio for $\phi$ $\rightarrow$
          $\tau$$\tau$ production as a function of $m_{\phi}$ assuming the SM width of the Higgs boson. 
Results (square marker) correspond to an integrated luminosity of $\mathcal{L}$=2.2~fb$^{-1}$ and
	            are compared to the D\O\ Run IIa $\mathcal{L}$=1.0~fb$^{-1}$ result (asterisk marker).
   }
   \label{fig:mvis}
\end{figure}

The $\Mvis$ spectrum 
shown in Fig.~\ref{fig:mvis}
is also used as the input to the limit calculator {\tt collie}~\cite{bib-wadelimit}
using the $CL_s$ method at 95\% Confidence Level.
These limits are shown in Fig.~\ref{fig:mvis} assuming a 
Higgs boson with SM width.  Correlations in systematic uncertainties between the different
tau decay channels that have been studied are taken into account.   The combination of the 2.2~fb$^{-1}$ 
data set analyzed at D\O\ in Run II provides a 10-20\% improvement in
the cross section across the range of Higgs boson masses studied
compared to the results from the 1.0~fb$^{-1}$ data set only.

\section{\label{sec:mssm}INTERPRETATION OF RESULTS IN MSSM}

Using the limits set on the cross section for neutral Higgs production,
regions of the ($m_{A}$, tan$\beta$) parameter space can be excluded in the
the MSSM.  Through radiative corrections beyond tree level, 
the masses and couplings of the Higgs boson depend on additional SUSY parameters.   
Assuming a CP-conserving Higgs sector, limits on tan$\beta$ as a function of
$m_{A}$ can be derived in two benchmark scenarios given by~\cite{bib-conf} and defined in~\cite{mssm-benchmark}:

\begin{itemize}
\item{\bf {\bf \boldmath{$m_{h}^{max}$}} scenario}
\item{\bf No-mixing scenario}
\end{itemize}

\begin{figure}[htbp]
   \begin{center}
     \subfigure[] {
       \includegraphics[width=7cm]{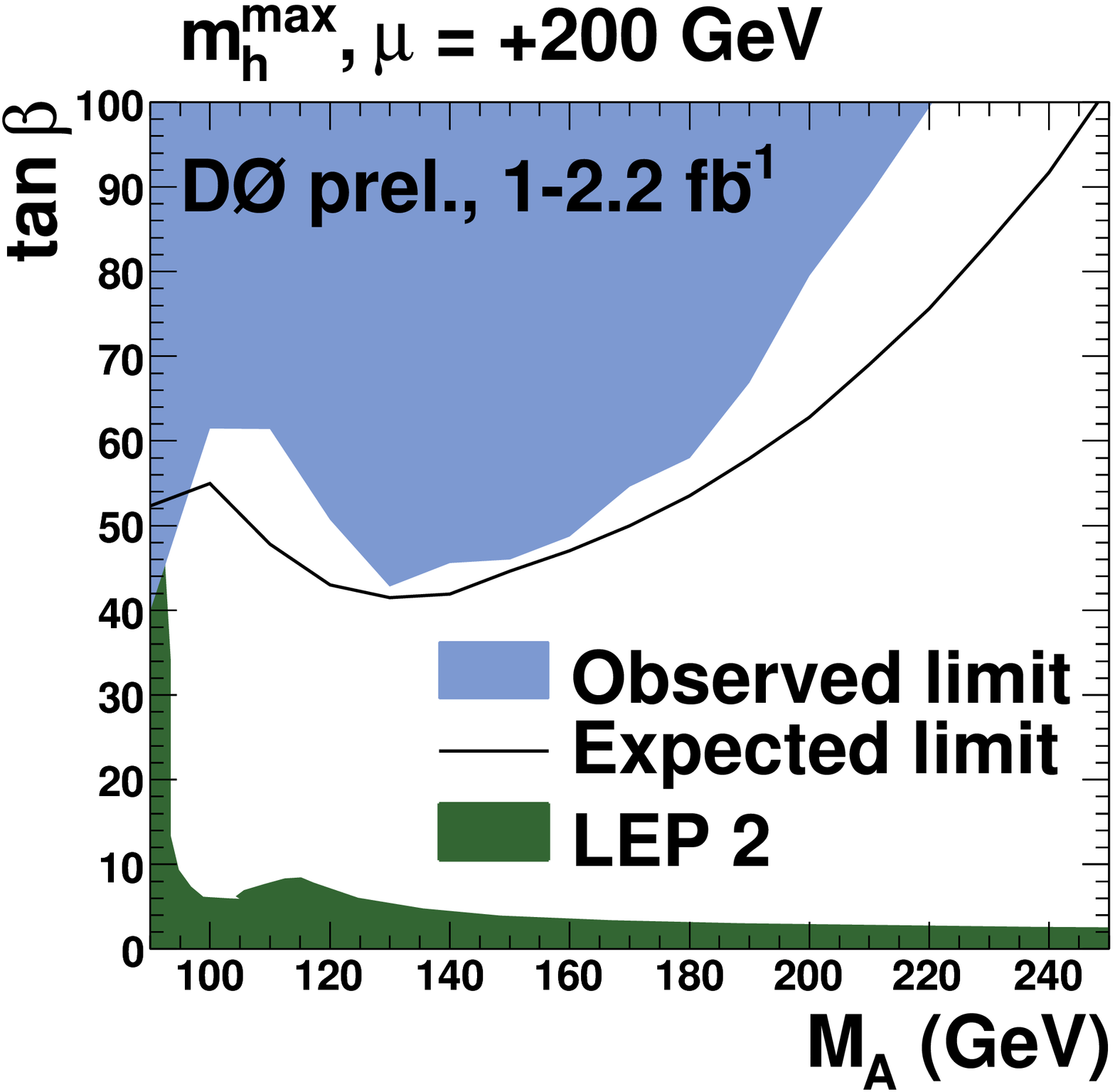}}
     \subfigure[] {
       \includegraphics[width=7cm]{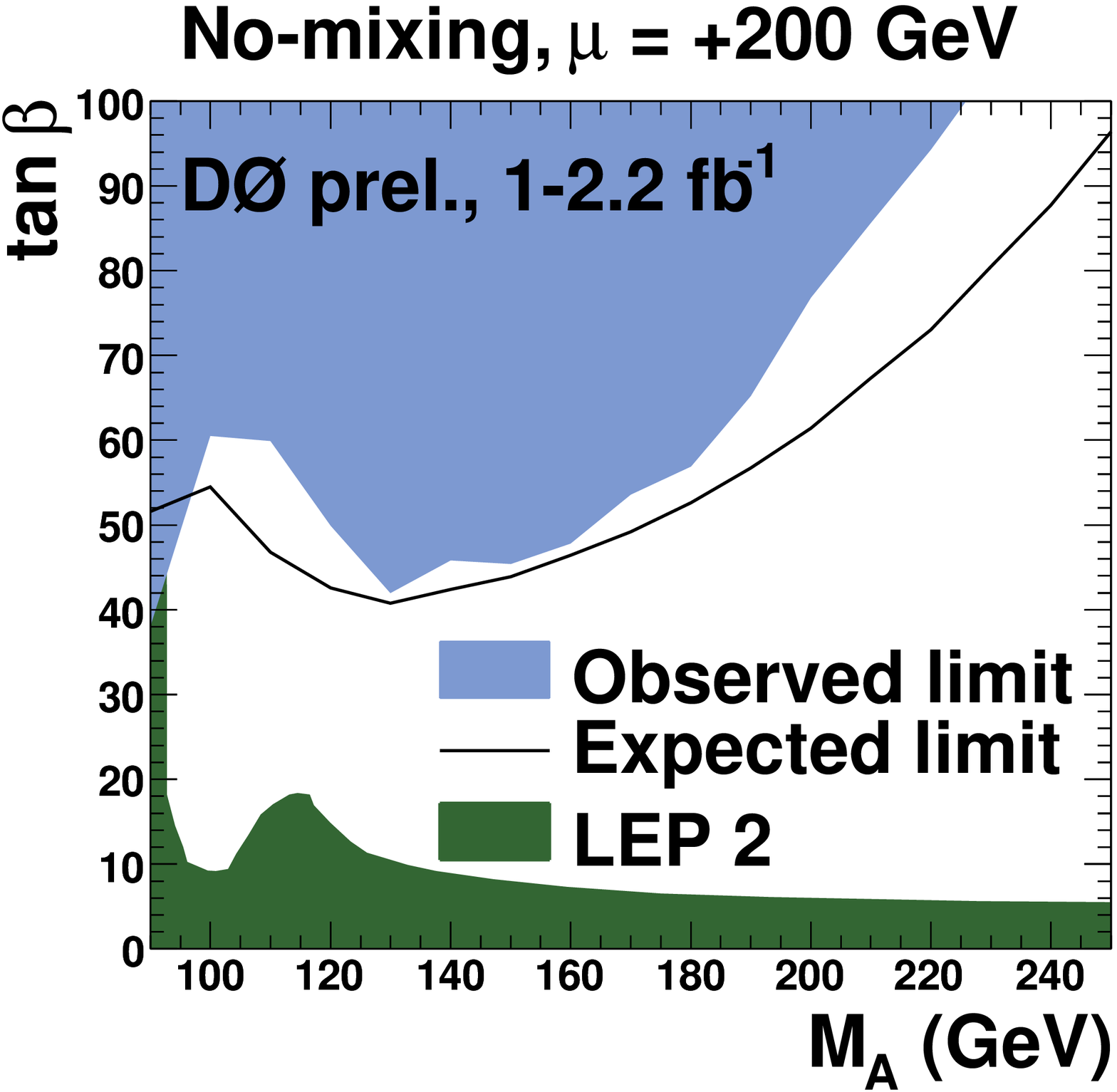}}
     \end{center}
   \caption{
    Region in the ($m_{A}$, tan$\beta$) parameter space that has been excluded at 95\% CL for $\mu >$ 0 in 
    two MSSM benchmark scenarios: (a) $m_{h}^{max}$ and (b) no-mixing.  
    Shown also by the green shaded region is the excluded region by LEP~\cite{bib-lep}.
   }
   \label{fig:mssm}
\end{figure}

The cross section, width, and branching ratios for the Higgs boson have been calculated using the {\tt FEYNHIGGS} program~\cite{feynhiggs} 
version 2.6.4.  
The excluded region in the ($m_{A}$, tan$\beta$) space in each scenario 
is given in Fig.~\ref{fig:mssm}, for the case of $\mu >$ 0.  The region excluded by the LEP experiments~\cite{bib-lep} is also shown.  The $\mu <$ 0 case, which is presently disfavored~\cite{bib-negmu}, is not considered.  
At large tan$\beta$, the $A$ boson is nearly degenerate in mass with either the $h$ or $H$ boson and thus,
the production cross sections for $gg \rightarrow \phi$ and $b$$\bar{b}$ $\rightarrow \phi$ are added at each 
($m_{A}$, tan$\beta$) point.  

\section{\label{sec:conclusion}CONCLUSION}
The search for the production of neutral Higgs bosons decaying
into $\tau^+\tau^-$ final states in the D\O\ Experiment is presented with $2.2$~fb$^{-1}$ integrated luminosity.
The result shows 25\% improvement in the Higgs cross-section limits around the mass range $200~GeV$
compared to the published result in $1.0$~fb$^{-1}$ data set, 
and a sensitivity to tan$\beta$ about 60 
for low $m_{A}$ bins and 40-50 for the rest of $m_{A} < 200~GeV$ has been reached. 



\end{document}